\documentclass[10pt]{article}
\usepackage{amsmath,graphicx,tabularx}
\usepackage{caption}

\usepackage{enumitem}
\setlist{nosep, leftmargin=14pt}

\usepackage{authblk}
\title{Automated Small Kidney Cancer Detection in Non-Contrast Computed Tomography}
\usepackage[a4paper, total={6in, 8in}]{geometry}
\begin{document}
\author{William McGough$^{*,1,2}$ Thomas Buddenkotte$^{3,4}$ Stephan Ursprung$^{2,5,6}$ \\ Zeyu Gao$^{1,2}$ Grant D. Stewart$^{2,6,7}$ Mireia Crispin-Ortuzar$^{1,2}$
\thanks{$ ^{\star}$ Corresponding Author: wcm23@cam.ac.uk}
}
\affil[1]{Department of Oncology, University of Cambridge}
\affil[2]{CRUK Cambridge Centre, University of Cambridge}
\affil[3]{Department of Diagnostic and Interventional Radiology and Nuclear Medicine, University Medical Center Hamburg-Eppendorf}
\affil[4]{jung diagnostics GmbH, Hamburg, Germany}
\affil[5]{Department of Radiology, University of Cambridge}
\affil[6]{Cambridge University Hospitals NHS Foundation Trust, Cambridge, UK}
\affil[7]{Department of Surgery, University of Cambridge}

\maketitle
\begin{abstract}
This study introduces an automated pipeline for renal cancer (RC) detection in non-contrast computed tomography (NCCT). In the development of our pipeline, we test three detections models: a shape model, a 2D-, and a 3D axial-sample model. Training (n=1348) and testing (n=64) data were gathered from open sources (KiTS23, Abdomen1k, CT-ORG) and Cambridge University Hospital (CUH). Results from cross-validation and testing revealed that the 2D axial sample model had the highest small ($\leq$40mm diameter) RC detection area under the curve (AUC) of 0.804. Our pipeline achieves 61.9\%  sensitivity and 92.7\% specificity for small kidney cancers  on unseen test data. Our results are much more accurate than previous attempts to automatically detect small renal cancers in NCCT, the most likely imaging modality for RC screening. This pipeline offers a promising advance that may enable screening for kidney cancers.
\end{abstract}

Cancer Detection, Segmentation, Deep Learning, Radiology
\section{Introduction}
\label{sec:intro}

RC is a significant global health concern, with 342,000 cases and 132,000 deaths recorded worldwide in 2016\cite{10.1001/jamaoncol.2018.2706}, and it is most commonly detected in CT scans of the abdomen. Detecting RC early improves survival rates\cite{TOSAKA19901097}, and previous studies have indicated the benefits of using NCCT for cancer screening. However, the high economic cost of radiological personnel limits the practicality of CT-based screening for diseases with less distinguishable risk factors, such as RC\cite{usher2020current}.  The development of an automated RC detection system would reduce the cost of diagnosis and reduce patient processing time, potentially enabling RC screening.

One of the key objectives of an automated cancer detection system, especially in the context of  screening, is the ability to identify small cancers. Detecting RC at a smaller size is associated with reduced mortality rates\cite{cheville2003comparisons}, yet, surprisingly, the ability of automated detection systems to detect smaller RCs in a suitable imaging mode for screening has received limited attention in existing research. Small RC detection via NCCT has previously involved the application of pretrained deep-learning classifiers to 2D axial samples from CT or MRI scans\cite{mcgough_sanchez_mccague_stewart_schönlieb_sala_crispin-ortuzar_2023, doi:10.2214/AJR.19.22074}. This approach works well in contrast-enhanced CT (CECT), where textural information is rich in the axial samples of CECT scans (AUC 0.846), but previous experiments have reported reduced accuracy in NCCT where textural information is degraded (AUC 0.562)\cite{doi:10.2214/AJR.19.22074}. Alternatively, it is possible to detect small RCs in CECT with a purely 3D segmentation approach (AUC 0.93)\cite{toda2022deep}. Elsewhere, lung cancer detection in NCCT is highly effective when using a model that exploits 3D information within the scan (AUC 0.959)\cite{ardila2019end}.

We present what we believe to be the first end-to-end automated RC detection system optimised for NCCT, the most likely imaging modality for RC screening. We use this system to show that detection of small renal cancers is possible using a 2D deep-learning image classifier in non-contrast CT, if the data preparation methodology is designed appropriately. We also test how 3D image classifiers and shape ensemble classifiers fare in small cancer detection in NCCT.

\section{Method}
\label{sec:format}
\begin{figure*}
  \centering
  \includegraphics[width=0.8\textwidth]{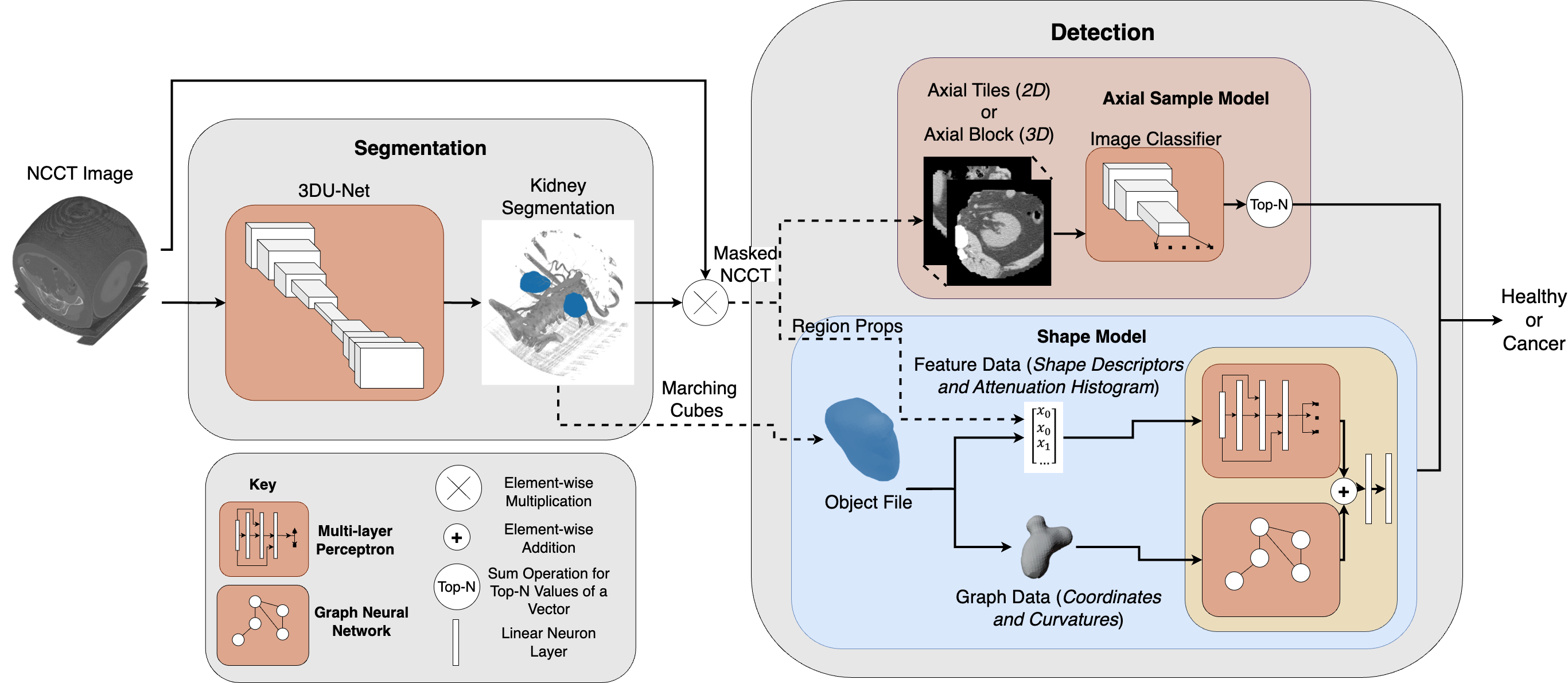}
  \caption{Our pipeline for automatic RC detection in NCCT. We show both the shape model and axial sample models occupying the pipeline's detection phase for illustrative purposes; in fully-automated inference, only one detection model is used at a time.}
  \label{fig:fig1}
\end{figure*}

Our end-to-end pipeline is shown in Fig. \ref{fig:fig1}; we describe each component of this pipeline in the following section.

\subsection{Segmentation and Object Generation} For automatic kidney segmentation, we used a 3DU-Net model - a deep-learning architecture specifically designed for volumetric segmentation tasks. We used a 6-stage 3DU-Net, with 32 3$^3$ convolutional filters in the first block, doubling sequentially until 1024 filters in the bottom block. Our 3DU-Net was trained to generated binary inferences; 1 for all tissue within the kidney contour (kidney, tumour, or cyst), and 0 for background. The 3DU-Net model was trained using the nnU-Net framework\cite{nnunet}. To generate object files for each kidney, we ran a marching cubes algorithm (using python package Scikit-Image) over the segmentation inference of each kidney, generating a set of vertices, edges, and faces. Then, these objects were sparsified and smoothed using Blender's Python API, using the remesh method with a voxel size 1.2, and smooth method with factor 0.5 and 5 iterations.

\subsection{Shape Ensemble} The shape ensemble aims to detect cancers purely using 3D spatial information extraced from the kidney object files. Our motivation for training a shape model was to enable the detection of exophytic tumours that we observed being missed by the 2D axial model. The shape ensemble consists of two distinct components: a multilayer perceptron (MLP) and a graph neural network (GNN). 

The GNN is a neural network that operates on graph-structured data. The GNN takes as input a graph representing the kidney's surface, which are generated using the vertices and edges of the kidney object. Each vertex in the graph contains a 4-element vector; 3 spatial coordinates and a value for curvature.  Curvature was calculated using the dot product of the difference between the neighbouring vertices' normal vectors and coordinate values, using the following equations

\begin{equation}\displaystyle C_{e,ij}= \frac{(\overrightarrow{n_i} - \overrightarrow{n_j}) \cdot \overrightarrow{v_i} - \overrightarrow{v_j}}{|\overrightarrow{v_i} - \overrightarrow{v_j}| + \epsilon} ,\end{equation}

\begin{equation}\displaystyle C_{v} = \frac{\sum_{j=1}^{N_i-1} \theta_{i,(j,j+1)} (C_{e,ij}+C_{e,ij+1})}{2\sum_{j=1}^{N_i-1} \theta_{i,(j,j+1)}},\end{equation}
where $C_{e,ij}$ is the curvature of the edge connecting nodes $i$ and $j$, $C_{v}$ is a node's curvature value, $\overrightarrow{n_i}$ is node $i$'s normal vector, $\overrightarrow{v_i}$ is node $i$'s spatial coordinate vector, $N_i$ is the number of neighbouring nodes to node $i$, $\theta_{i,(j,j+1)}$ is the angle between neighbouring nodes $j$ and $j+1$ from the perspective of a reference node $i$, and $\epsilon$ is a small fixed value; we used $10^{-6}$. The GNN is a sequential 5-layer graph convolutional neural network, using Chebyshev convolutional layers with a filter width of 2 nodes and 25 dimensional latent feature space.

The MLP is a 3-layer fully connected neural network. Its input data are: a set of shape descriptors (volume, maximum and minimum diameter, convexity; the three eigenvalues from the inertia tensor of the segmented volume, and a binary feature to indicate kidney location), a histogram of curvatures (taken from the object graph, 10 bins between [-0.5,0.5]), and a histogram of kidney attenuations (10 bins between [-20,80] Hounsfield Units (HU)), totalling 28 input features. All shape descriptors were extracted using the regionprops function from Scikit-Image. To fuse the shape ensemble, we replace both models' final classification layer with a linear layer that projects each model's features into a common latent space. These features are added and fed into a new, ensemble-wide classification layer, which allowed the whole ensemble to be trained end-to-end.

\subsection{Axial Sample Classifier} An axial sample model is a deep-learning image classifier, already implemented in torchvision, retrained to identify RC from axial CT samples. Two versions of the axial sample classifier were assessed: 2D and 3D.  Both were pretrained on KiTS23 CECT data and fine-tuned on a dataset of NCCT samples. In 2D, these samples were of size 1x224x224 - we call these `tiles'. In 3D, samples were of size 20x224x224 - we call these `blocks'. We jointly refer to tiles and blocks as `samples' where the distinction is unimportant. 

Following benchmarking (see Table \ref{table:AUCtable}), we selected the top-2 architectures (EfficientNet and ResNeXt) as backbones for our axial sample models. Both architectures were expanded to 3D by converting all 2D PyTorch layers to 3D via symmetric extension of all properties; for example, (3x3) kernels become (3x3x3). Our 3D EfficientNet model was based on EfficientNet V2-L, but our 3D ResNeXt model was based upon the smaller ResNeXt-50 32x8d, as expanding ResNeXt to 3D increases parameter count roughly 50 times. To establish kidney-wise cancer from sample inferences, we employed a sample voting method. Specifically, we calculated the sum probability score across the most-probable samples for each kidney. In 2D, we used the top-10 most probable tile scores. In 3D, we used the top-1 most probable block score. 
\section{Experiments}
\subsection{Datasets} This research study used open access segmented data (KiTS23, Abdomen1k, CT-ORG) and data made available from CUH. Ethical approval was only required for CUH data, and it was gathered for all patients. Sample data were generated from both the KiTS23 dataset and CUH data; KiTS23 contains 489 CECT scans, and CUH data contains 92 paired CECT and NCCT scans. Segmentations were generated for all CUH CECT scans via a segmentation 3DU-Net and a radiologist that labelled renal tumours accurately. A subset of KiTS23 cases (110 cases) had paired NCCT scans, sourced from the Climb 4 Kidney Cancer (C4RC) dataset. We coregistered 79 NCCT scans to their equivalent CECT in KiTS23, and 45 from our CUH dataset, generating 124 labelled NCCT scans - this is referred to as the coregistered dataset (CD).  The remaining unlabelled 47 NCCT scans from CUH are combined with 17 NCCT non-cancerous cases (7 from Abdomen1K, 10 from CT-ORG), sourced from open-source datasets, to form a test set.

For segmentation training, scans were converted into an isotropic voxel size of 4mm, clipped to 0.05-99.5 percentile attenuation values, and attenuation-normalised. To generate axial samples, we used 122 NCCT scans from the CD; two patients with `horseshoe' kidneys were excluded. Before sampling, each scan was: converted into an isotropic voxel size of 1mm, attenuation-clipped between [-200, 200] HU, normalised by dividing attenuation values by 100, and masked by the kidney segmentation (dilated by 4cm). 

Samples were taken from CT images using two schemes: centralised and sliding-window. Centralised samples were central upon the kidney's axial-plane centroid. Sliding-window samples were sampled uniformly across the scans with a 40x40mm axial-plane spacing. In both schemes, a 5mm out-of-plane spacing was used in 3D, and 1mm in 2D. Samples were assigned a `cancerous' label if their respective segmentation contained a cancer region larger than 1cm radius; else, `normal kidney' if it contained a kidney region larger than 2cm radius; otherwise, a `none' label.  A maximum of 50 sliding-window samples were taken per-class, per-kidney. From the CD, 21,953 normal kidney, 10,296 cancerous, and 28,186 none 2D tiles were generated, and 10,286 normal kidney, 7520 cancerous, and 9270 none 3D blocks were generated. From KiTS23, 73,167 normal kidney, 33,635 cancerous, and 101,872 none 2D CECT tiles were generated, and 34,156 normal kidney, 25,294 cancerous, and 32,383 none 3D CECT blocks were generated.

1011 kidney contours were sourced from KiTS23 and CD for the shape models to train upon---we call this the shape dataset. All shape data were given a binary label, either positive (distorted contour), or negative (normal). In training, we avoided the cancer/non-cancer dichotomisation due to cysts in the training data; shape models are na\"ive to the cause of contour perturbation. We used label thresholding for the shape data to clarify the training signal and improve model performance; if the kidney contours contained a cancer or cyst above 500mm$^3$, their graph label was positive, and if above 20000mm$^3$, their MLP feature set label was positive, too. When training the shape ensemble, we used a common threshold of 500mm$^3$. Label thresholds were selected in a hyperparameter search maximising cross-validation AUC. 

\subsection{Training Overview}
We trained all models in patient-wise 5-fold cross-validations, using the Adam optimiser and cross-entropy loss function. All training was performed on University of Cambridge's Wilkes3 HPC and University of Birmingham's Baskerville HPC cluster, on NVIDIA A100 GPUs, using CUDA-backend PyTorch v11.7, torchvision 0.15.2, and DGL 1.0.2 for GNNs.
\begin{table*}
\centering
\resizebox{\textwidth}{!}{
\begin{tabular}{|c|c|c|c|c|c|c|}
\hline
\textbf{Model Name} & \textbf{ResNet-152} & \textbf{ResNeXt-101 32x8d} & \textbf{ConvNeXt-B} & \textbf{EfficientNet-V2-L} & \textbf{Swin-V2-B} & \textbf{ViT-B-32} \\
\hline
\textbf{Case-wise AUC} & 0.810 & 0.827 & 0.807 & \textbf{0.837} & 0.596 & 0.752 \\
\hline
\end{tabular}}
\caption{Results from benchmarking on the CD 2D tile dataset. Architecture versions were selected by matching parameter count. All models were pretrained on ImageNet-1k. Reported AUCs are the mean of 3 five-fold patient-wise cross-validations.}
\label{table:AUCtable}
\end{table*}
\subsubsection{Segmentation}
We pretrained 3DU-Net on the combination of KiTS23, Abdomen1k and CT-ORG CECT data (n=1303), removing NCCT cases, before fine-tuning on the CD. We trained with a learning rate of 10$^{-4}$, rising to 4x10$^{-3}$ for pretraining and 5x10$^{-4}$ for fine-tuning, with 500 epochs for pretraining and 100 epochs for fine-tuning.  We used random rotations, flips, contrast- and blur-adjustment for data augmentation. We used a batch size of 16; training inputs were randomly sampled 64$^3$ voxel cubes; epoch length was 250 batches. We used a linear-ascent exponential-decay schedule:

\begin{equation}\displaystyle LR(k) = \begin{cases} 
      LR_{max}(\frac{k-1}{15}) + LR_{min} & k\leq 15 \\
      LR_{max}(e^{-a\frac{k-16}{k_{max}-k}}) & k > 15
   \end{cases}
\end{equation}
where $LR(k)$ is learning rate at epoch $k$, and $a$ is a decay constant;  $a=4$ for pretraining and $a=3$ for fine-tuning.

\subsubsection{Shape Ensemble and Axial Sample Models}
 Both shape models were trained individually for 100 epochs; MLP learning rate was 10$^{-2}$, GNN learning rate was 10$^{-3}$. After, we froze the individual models, trained the ensemble's shared layers for 30 epochs, before training all layers for 2 epochs, with a learning rate of 10$^{-3}$ and batch size of 8. We pretrained both axial models on KiTS23 CECT data (5 epochs, learning rate of 1x10$^{-3}$) before finetuning on CD (5 epochs, learning rate of 5x10$^{-4}$). We used a batch size of 16 and the augmentation strategy from segmentation training. 
\begin{figure}[t]
  \centering
  \includegraphics[width=0.88\linewidth]{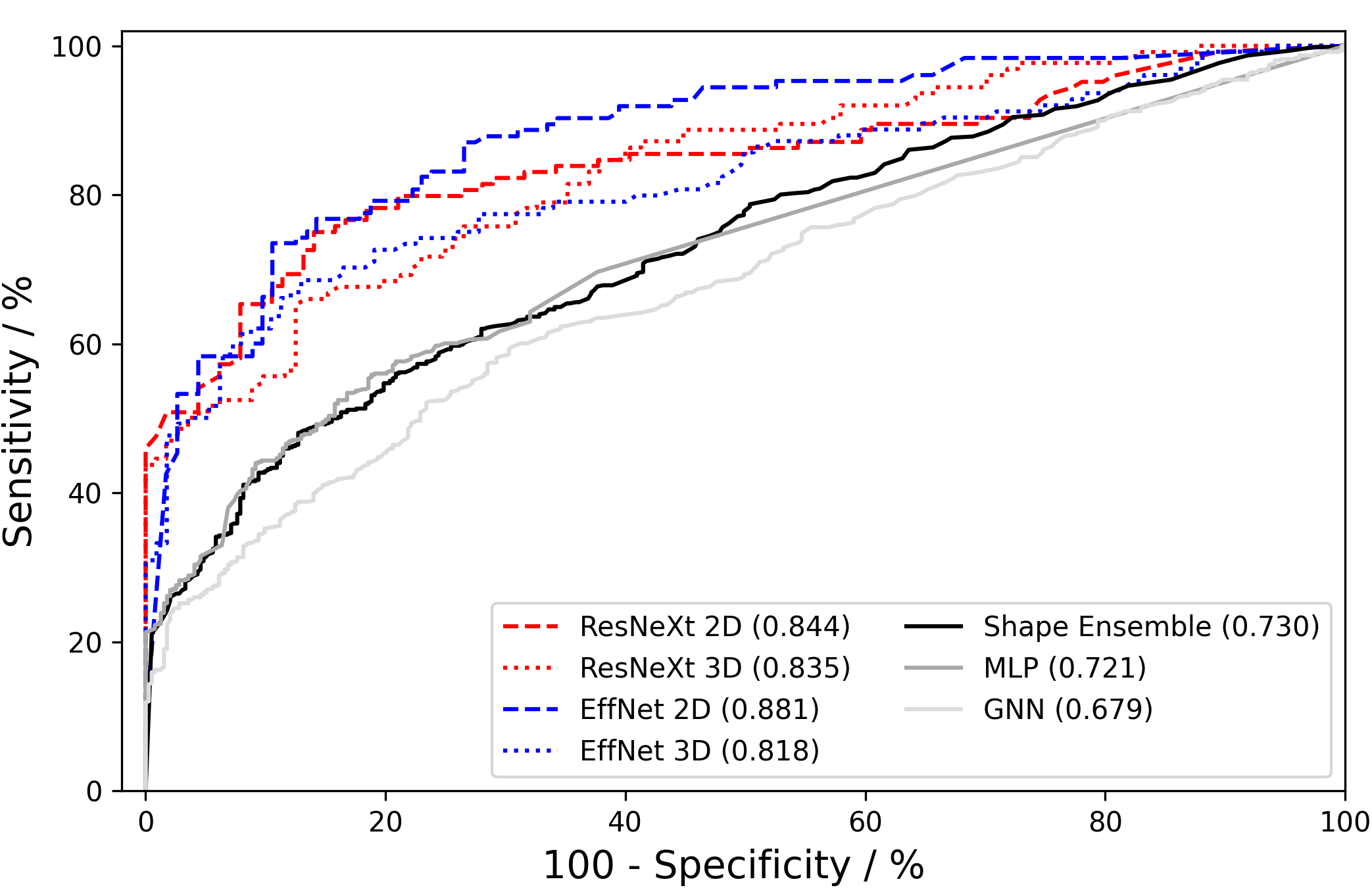}
  \caption[width=7.5cm]{Cross-validation performance for all models. Sample models were evaluated over the CD following pretraining; shape models were evaluated over the shape dataset.}
  \label{fig:ROC}
\end{figure}
\subsection{Testing}
 In our test data, 47 patients had unilateral RC and 17 had healthy kidneys. For each model (Shape, 2D, 3D), we calculated per-kidney detection sensitivity and specificity and plot its receiver operating characteristic (ROC); for each model type, inference was performed by summing the output from the 5 cross-validation models. In axial sample models, inference was only applied to `centralised' axial samples that contain a kidney region from the segmentation inference. 

\begin{figure}[t]
  \centering
  \includegraphics[width=0.90\linewidth]{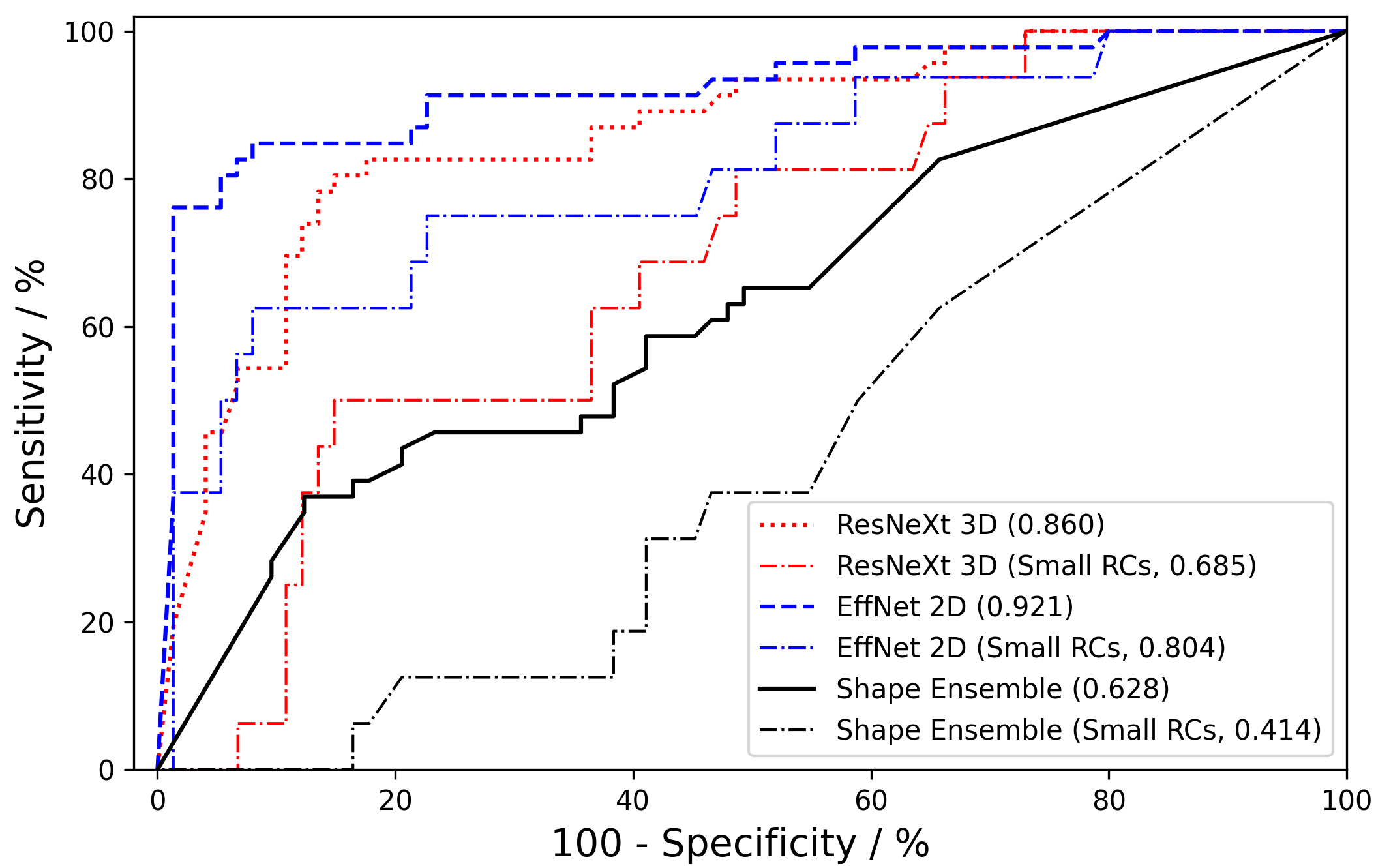}
    \caption[width=7.5cm]{Test performance of select models from cross-validation on unseen test data (n=65).}
 \label{fig:test}
\end{figure}
\section{Results and Discussion}
The cross-validation Dice Similarity Coefficient of 3DU-Net was 90.13. Fig. \ref{fig:ROC} shows the cross-validation performance of each detection model; 2D EfficientNet had the highest AUC (0.881). Fig. \ref{fig:ROC} shows that all axial models perform similarly at high specificities, and the shape ensemble performs better than either the MLP or GNN in isolation, validating the ensemble. Fig. \ref{fig:test} shows the test performance of each detection model; RC detection sensitivity by all models is highly dependent upon size, with models detecting small RCs with lower accuracy. Fig. \ref{fig:test} shows that cancer detection via the shape ensemble is possible, but that both axial models considerably outperform this method, especially for small RCs. The 2D model had the highest test AUC (0.921) and detected small unseen RCs with 61.9\% sensitivity and 92.7\% specificity.

Possible reasons for the superiority of the 2D approach over the 3D approach are: a greater volume of training samples (6.04x10$^4$ NCCT tiles; 2.70x10$^4$ NCCT blocks), a higher labelling density in 2D (one label per 5x10$^4$ pixels in 2D; 1 label per 1x10$^6$ voxels in 3D), and the availability of ImageNet pretraining in 2D. The effect of many of these advantages would be reduced with a larger 3D dataset. Possible reasons for the superiority of 2D model over previous attempts\cite{doi:10.2214/AJR.19.22074} are: an improved deep-learning architecture (EfficientNet vs Inception), pretraining in CECT, improved sampling scheme (using sliding-window and centralised samples over kidney), and using nnU-Net's augmentation techniques. We note that our models had access to less training NCCT scans (n=124) than this previous attempt (n=168).
\section{Conclusion}
We present an automated RC classification pipeline optimised for NCCT and provide evidence of its efficacy in detecting small RCs. Our 2D tile model (AUC 0.804) far exceeds previous attempts at detecting small RCs automatically in NCCT (AUC 0.562), and may enable RC screening via NCCT.
\section{Acknowledgments}
\label{sec:acknowledgments}

The Baskerville HPC system that made this research possible is funded by the EPSRC and UKRI through the World Class Labs scheme (EP/T022221/1) and the Digital Research Infrastructure programme (EP/W032244/1) and is operated by Advanced Research Computing at the University of Birmingham. This work was supported by the Cancer Research UK Cambridge Centre [CTRQQR-2021/100012] and the International Alliance for Cancer Early Detection, a partnership between Cancer Research UK [C14478/A27855], Canary Center at Stanford University, the University of Cambridge, OHSU Knight Cancer Institute, University College London and the University of Manchester. This work was supported by The Mark Foundation for Cancer Research [RG95043], the Cancer Research UK Cambridge Centre [C9685/A25177 and CTRQQR-2021/100012] and NIHR Cambridge Biomedical Research Centre (NIHR203312). The views expressed are those of the authors and not necessarily those of the NHS, the NIHR or the Department of Health and Social Care. 

 TB is a full-time employee at jung diagnostics. GDS has received educational grants from Pfizer, AstraZeneca and Intuitive Surgical; consultancy fees from Pfizer, MSD, EUSA Pharma and CMR Surgical; Travel expenses from MSD and Pfizer; Speaker fees from Pfizer; Clinical lead (urology) of National Kidney Cancer Audit and Topic Advisor for the NICE kidney cancer guideline. MCO is a co-founder and CDO of 52 North Health.

\section{Compliance with Ethical Standards}
This research study was conducted retrospectively using human subject data shared via open source and other data shared privately for research purposes. This study was performed in line with the principles of the Declaration of Helsinki. Approval was granted for use of the private research data by the Human Biology Research Ethics Committee of University of Cambridge, on 25 April 2023.

\bibliographystyle{IEEEbib}
\bibliography{strings,refs}

\end{document}